\documentclass[epj]{webofc}
\usepackage[utf8]{inputenc}
\usepackage[varg]{txfonts}   
\usepackage{booktabs}
\usepackage{xcolor}
\definecolor{darkred}{rgb}{0.4,0.0,0.0}
\definecolor{darkgreen}{rgb}{0.0,0.4,0.0}
\definecolor{darkblue}{rgb}{0.0,0.0,0.4}
\usepackage[bookmarks,linktocpage,colorlinks,
    linkcolor = darkred,
    urlcolor  = darkblue,
    citecolor = darkgreen]{hyperref}
\newcommand {\beq} {\begin{equation}}
\newcommand {\eeq} {\end{equation}}
\newcommand {\beqa}{\begin{eqnarray}}
\newcommand {\eeqa}{\end{eqnarray}}

\usepackage{enumitem}
\usepackage{subfigure}
\wocname{EPJ Web of Conferences}
\woctitle{Lattice2017}
\begin{document}

\begin{flushright} 
KEK-TH-2008
\end{flushright} 

\vspace{-0.3cm}

\selectlanguage{english}
\title{%
Unification of the complex Langevin method and
the Lefschetz-thimble method
}
\author{%
\firstname{Jun} \lastname{Nishimura}\inst{1,2}\fnsep\thanks{Speaker, \email{jnishi@post.kek.jp}} \and
\firstname{Shinji} \lastname{Shimasaki}\inst{1,3}
}
\institute{%
KEK Theory Center, High Energy Accelerator Research Organization,\\
1-1 Oho, Tsukuba, Ibaraki 305-0801, Japan
\and
Graduate University for Advanced Studies (SOKENDAI),\\
1-1 Oho, Tsukuba, Ibaraki 305-0801, Japan
\and
Research and Education Center for Natural Sciences, Keio University,\\
Hiyoshi 4-1-1, Yokohama, Kanagawa 223-8521, Japan
}
\abstract{%
Recently there has been remarkable progress in solving the sign problem, 
which occurs in investigating statistical systems with a complex weight. 
The two promising methods, the complex Langevin method and the Lefschetz 
thimble method, share the idea of complexifying the dynamical variables, 
but their relationship has not been clear. Here we propose a unified 
formulation, in which the sign problem is taken care of by both the 
Langevin dynamics and the holomorphic gradient flow. We apply our 
formulation to a simple model in three different ways and show that one 
of them interpolates the two methods by changing the flow time.
}
\maketitle
\section{Introduction}\label{intro}

Recently there has been remarkable development 
in solving the sign problem.
Suppose we consider a path integral
\beq
Z=\int dx \, w(x) \ ,
\eeq
in which the integration contour
is taken to be the real axis and the integrand $w(x)$
is a complex number.
When the phase of $w(x)$ oscillates violently,
it is extremely hard to evaluate the integral by Monte Carlo methods
using the reweighting formula due to huge cancellations between
configurations sampled with the Boltzmann weight $|w(x)|$.
This is the sign problem.
The key idea in recent development
is to complexify the dynamical variables.

In the generalized Lefschetz-thimble method (GLTM)
\cite{Cristoforetti:2012su,Alexandru:2015sua},
the sign problem is minimized by deforming the integration contour.
Namely, the integration contour is determined by
solving the holomorphic gradient flow, which
makes the phase oscillation of the integrand much milder.
The remaining phase oscillation can be dealt 
with by the standard reweighting method.
In the complex Langevin method (CLM) \cite{Parisi:1984cs,Klauder:1983sp}, 
on the other hand,
one defines a stochastic process in the complex plane
in such a way that the ensemble average using configurations
after thermalization gives the expectation value
with respect to the original path integral.
Unlike the GLTM, there is no remaining phase to take care of.
However, the equivalence to the original path integral
requires certain conditions \cite{Aarts:2009uq,Aarts:2011ax,Nagata:2016vkn},
which limit the applicability of the method.
Here we propose a unified formulation \cite{Nishimura:2017vav},
which clarifies the relationship between the two methods.

The rest of this article is organized as follows.
In section \ref{sec:review}
we briefly review the CLM and the GLTM.
In section \ref{sec:unify} we present our formulation,
which unifies the two methods.
In section \ref{sec:0d-model}
we apply our formulation to
a single-variable case and
clarify the relationship between the two methods.
Section \ref{sec:summary}
is devoted to a summary and discussions.

\section{Brief review of the CLM and the GLTM}
\label{sec:review}

In this section, we briefly review the CLM and the GLTM.
Here we consider a general model given by a multi-variable integral 
\begin{align}
Z=\int_{\mathbb{R}^n} dx \, e^{-S(x)} \ ,
\label{Z}
\end{align}
where the action $S(x)$ is a complex function of 
$x=(x_1,\cdots,x_n) \in \mathbb{R}^n$.
The expectation value of an observable $\mathcal O(x)$ is defined by
\begin{align}
\langle \mathcal O(x)\rangle =
\frac{1}{Z}\int_{\mathbb{R}^n} dx \, \mathcal O(x) \, e^{-S(x)} \ .
\label{O}
\end{align}

\subsection{Complex Langevin Method (CLM)}

In the CLM \cite{Parisi:1984cs,Klauder:1983sp}, 
we complexify the dynamical variables 
$x\in\mathbb{R}^n$ as $x\to z= x+iy\in \mathbb{C}^n$
and extend the action and the observable to holomorphic functions of $z$.
Then we consider the complex Langevin equation
\begin{align}
\frac{\partial}{\partial t}z_k(t)
&=-\frac{\partial S(z)}{\partial z_k}+\eta_k(t) \ ,
\label{cle}
\end{align}
where $t$ is a fictitious time and 
$\eta_k(t)$ is a real Gaussian noise normalized as 
$\langle \eta_k(t)\eta_l(t') \rangle_\eta = 2 \, \delta_{kl} \, \delta(t-t')$.
This equation (\ref{cle}) defines a probability distribution of 
$z$ at each time $t$.
It is shown \cite{Aarts:2009uq,Aarts:2011ax} that under certain conditions,
the expectation value $\langle\mathcal O(x)\rangle$
in the original integral \eqref{O}
can be obtained as
\begin{align}
\langle\mathcal O(x)\rangle = \langle \mathcal O(z)\rangle_{\mathrm{CLM}} \ ,
\label{clm ob}
\end{align}
where the right-hand side
denotes the expectation value of $\mathcal O(z)$ 
with respect to the limiting distribution of $z$ at $t\to \infty$.
As in the case of Monte Carlo methods,
$\langle \mathcal O(z)\rangle_{\mathrm{CLM}}$
can be replaced with a long-time average of the observable
calculated for the generated configurations $z(t)$
if ergodicity holds for the Langevin time-evolution.
When the action $S(x)$ is real, this method reduces to the 
(real) Langevin method, which is a standard Monte Carlo algorithm.

The derivation of the equality \eqref{clm ob}
uses integration by parts, which can be justified
if the distribution of $z$ falls off fast enough
in the imaginary directions \cite{Aarts:2009uq,Aarts:2011ax}
as well as near the singularities of the drift term 
if they exist \cite{Nishimura:2015pba,Aarts:2017vrv}.
Recently \cite{Nagata:2016vkn}, a subtlety in the use of time-evolved 
observables in the original argument \cite{Aarts:2009uq,Aarts:2011ax} 
was pointed out, and the derivation of the equality \eqref{clm ob}
has been refined taking account of this subtlety.
This also led to the proposal of a useful criterion for justification,
which states that
the distribution of the drift term should 
be suppressed exponentially or faster at large magnitude \cite{Nagata:2016vkn}.

\subsection{Generalized Lefschetz-Thimble Method (GLTM)}

In the GLTM \cite{Alexandru:2015sua}, 
we deform the integration contour from $\mathbb{R}^n$ 
to an $n$-dimensional real manifold in $\mathbb{C}^n$
by using the so-called holomorphic gradient flow,
which makes the sign problem milder.

The holomorphic gradient flow is defined by
the differential equation 
%
\begin{align}
\frac{\partial}{\partial \sigma}\phi_k(x;\sigma) 
= \overline{ \left(\frac{\partial S(\phi(x;\sigma))}{\partial \phi_k}\right)} \ ,
\label{hge}
\end{align}
which is solved from $\sigma = 0$ to $\sigma=\tau$
with the initial condition $\phi(x;0)=x\in \mathbb{R}^n$.
The flowed configurations define
a $n$-dimensional real manifold in $\mathbb{C}^n$,
which we denote as $M_\tau=\{\phi(x;\tau)|x\in\mathbb{R}^n\}$.
One can actually argue that
the integration contour can be deformed continuously
from $\mathbb{R}^n$ to $M_\tau\subset \mathbb{C}^n$
without changing the partition function.
Then, by noting that 
$\phi(x)\equiv\phi(x;\tau)$
defines a one-to-one map
from $x\in \mathbb{R}^n$ to $\phi\in M_\tau$,
one can rewrite the partition function as
\begin{align}
Z
=\int_{M_\tau} d\phi \, e^{-S(\phi)}
=\int_{\mathbb{R}^n} dx \, \mathrm{det}J(x) \, e^{-S(\phi(x))} \ ,
\label{gltmZ}
\end{align}
where $J_{kl}(x)$
is the Jacobian corresponding to the map $\phi(x)$.
Thus, the deformation of the integration contour
simply amounts to changing the action from $S(x)$ to the ``effective action''
\begin{align}
S_{\mathrm{eff}}(x)=S(\phi(x)) - \log \mathrm{det} J(x) \ .
\label{Seff}
\end{align}

The virtue of using the holomorphic gradient flow
\eqref{hge} in deforming the integration contour
lies in the fact that
the real part of the action $S$ grows monotonically along the flow
keeping the imaginary part constant.
Thus, for a sufficiently large flow time $\tau$,
the partition function \eqref{gltmZ} is dominated by a small region of $x$,
and the sign problem becomes much milder.
One can therefore apply a standard Monte Carlo method
using only the real part of the effective action \eqref{Seff},
dealing with the imaginary part
by reweighting.
Thus, the expectation value of $\mathcal O(x)$ can be calculated as
\begin{align}
\langle \mathcal O(x)\rangle
=\frac{\langle e^{-i\mathrm{Im}S_{\mathrm{eff}}(x)}
\mathcal O(\phi(x;\tau))\rangle_{\mathrm{MC}}}
{\langle e^{-i\mathrm{Im}S_{\mathrm{eff}}(x)}\rangle_{\mathrm{MC}}} \ ,
\label{gltm ob}
\end{align}
where $\langle\cdots\rangle_{\mathrm{MC}}$ represents 
the expectation value obtained by a Monte Carlo method
with the weight $\exp(-\mathrm{Re}S_{\mathrm{eff}}(x))$.

When the flow time $\tau$ becomes infinitely large, $M_\tau$ contracts 
to a set of Lefschetz thimbles,
and the GLTM reduces to the so-called 
Lefschetz-thimble method \cite{Cristoforetti:2012su}.
Since $\mathrm{Im}S(\phi(x))$ is constant 
on each thimble, the sign problem is maximally reduced in some sense.
However, 
when there are more than one thimbles,
the transition between thimbles does not occur 
during the Monte Carlo simulation,
which leads to the violation of 
ergodicity.
The GLTM \cite{Alexandru:2015sua}
avoids this problem of the original method
by choosing a large but finite flow time.
Recently, it has been pointed out \cite{Fukuma:2017fjq,Alexandru:2017oyw} 
that the flow time can be made as large as one wishes
without the ergodicity problem 
if one uses the parallel tempering algorithm.

Let us here compare the two methods.
From the viewpoint of solving the sign problem,
the CLM is more powerful since it solves it completely as far as
it works, whereas the GLTM has a residual sign problem, which
may be problematic when the system size becomes large.
From the viewpoint of numerical cost, 
the GLTM is more demanding since one has to calculate the
Jacobian arising from deforming the integration contour.
It should also be noted that
one can include fermions in the CLM
with the cost of O($V$) for the system size $V$, whereas
in the GLTM, it requires O($V^3$).
The disadvantage of 
the CLM is that 
there is a 
condition \cite{Aarts:2009uq,Aarts:2011ax,Nagata:2016vkn}
to be satisfied to make it work, and hence there is a certain
range of applicability.

In this situation, it is useful to clarify the relationship
between the two methods.
Note here that despite the resemblance of the two methods in that they 
both use complexified variables, 
the connection between them is far from obvious.
In particular, the CLM samples configurations in the 
complex plane based on the Langevin process,
whereas the GLTM samples configurations on a particular contour
in the complex plane
obtained by deforming the original contour along the real axis.
Therefore the dimension of the configuration space 
in the CLM is twice as large as that of the GLTM.

\section{Unifying the CLM and the GLTM}
\label{sec:unify}

In this section, we present our formulation, which unifies
the CLM and the GLTM.
The basic idea is to 
apply the CLM to the partition function \eqref{gltmZ}
that appears in the GLTM.
The real variables $x\in \mathbb{R}^n$ 
parametrizing the $n$-dimensional real manifold in $\mathbb{C}^n$
have to be complexified, and they evolve in time following the
complex Langevin equation obtained from the effective action (\ref{Seff}).
The observable should be evaluated with 
the flowed configuration $\phi(z)$, which is defined 
as holomorphic extension of $\phi(x)$.
Therefore, it is important to see
how the distribution of flowed configurations
changes as the flow time increases.




We apply the CLM to the partition function \eqref{gltmZ}
by considering
the drift term
\begin{align}
\frac{\partial S_{\mathrm{eff}}(x)}{\partial x_k}=
\frac{\partial S(\phi(x))}{\partial \phi_l}J_{lk}(x)
- J^{-1}_{lm}(x)K_{mlk}(x) \ ,
\label{comb drift x}
\end{align}
where we have defined 
$K_{mlk}(x)\equiv \partial J_{ml}(x;\tau)/\partial x_k$.
In order to evaluate this function at complex values $z$,
we need to define
$\phi(z)$, $J_{kl}(z)$ and $K_{mlk}(z)$
as holomorphic extension of $\phi(x)$, $J_{kl}(x)$ and $K_{mlk}(x)$, 
respectively.

Let us note here that
the flow equation \eqref{hge}
for $\phi(x)$ involves complex conjugation on the right-hand side.
Therefore, in order to define
$\phi(z;\sigma)$
as holomorphic functions of $z$,
we have to extend the flow equations as
\begin{align}
\frac{\partial}{\partial \sigma}\phi_k(z;\sigma) 
&= \overline{ \left(\frac{\partial S(\phi(\bar z;\sigma))}{\partial \phi_k}\right)} \ ,
\label{hol hge-phi}
\end{align}
where
$\phi(\bar z;\sigma)$ appears on the right-hand side.
This function with $\bar z$ in the argument
obeys the flow equation, which can be obtained by replacing
$z$ with $\bar z$ in the above equation.
In practice, we solve the two sets of equations simultaneously
with the initial conditions 
$\phi(z;0)=z$
for a particular value of $z$ at each Langevin step.
The same trick can be used to obtain
$J_{kl}(z;\sigma)$ and $K_{mlk}(z;\sigma)$ as well.

Using this formulation,
we can calculate
the expectation value 
\eqref{O} as
\begin{align}
\langle\mathcal O(x)\rangle 
= \langle \mathcal O(\phi(z;\tau))\rangle_{\mathrm{CLM}} \ ,
\end{align}
where the flowed configuration $\phi(z;\tau)$ has to be used
in evaluating the observable unlike in the ordinary CLM \eqref{clm ob}.
For $\tau=0$, our formulation reduces to the ordinary CLM
since $\phi(z;0)=z$, $J_{kl}(z;0)=\delta_{kl}$ and $K_{klm}(z;0)=0$.


\section{Results for a single-variable model}
\label{sec:0d-model}

In this section, we apply our unified formulation
to a single-variable model to clarify
the relationship between the CLM and the GLTM.
The model is defined by the partition function
\cite{Nishimura:2015pba,Nagata:2016vkn,Aarts:2017vrv},
\begin{align}
Z=\int dx \, (x+i\alpha)^p \, e^{-x^2/2} \ ,
\label{single variable Z}
\end{align}
where $x$ is a real variable, 
while $\alpha$ and $p$ are real parameters. 
When $\alpha \neq 0$ and $p\neq 0$, the weight becomes complex and
the sign problem occurs.

\begin{figure}[t]
\centering
\includegraphics[width=7cm]{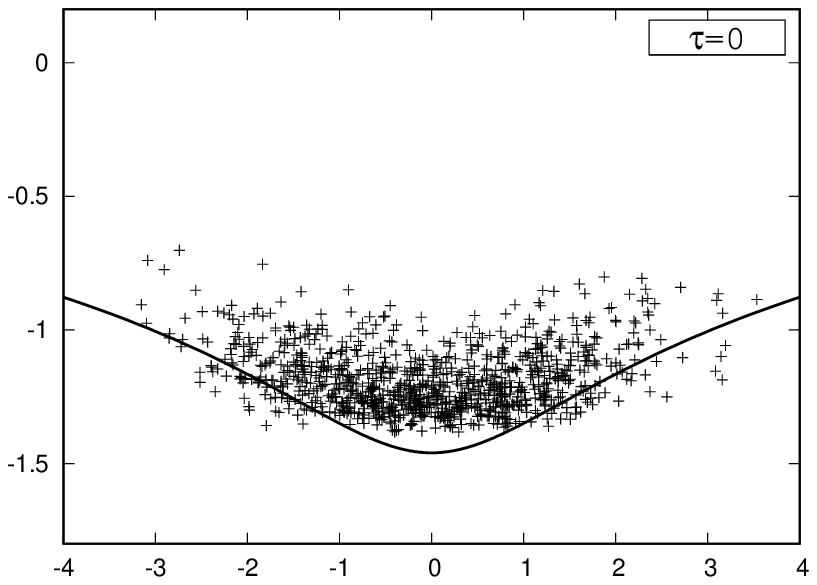}
\includegraphics[width=7cm]{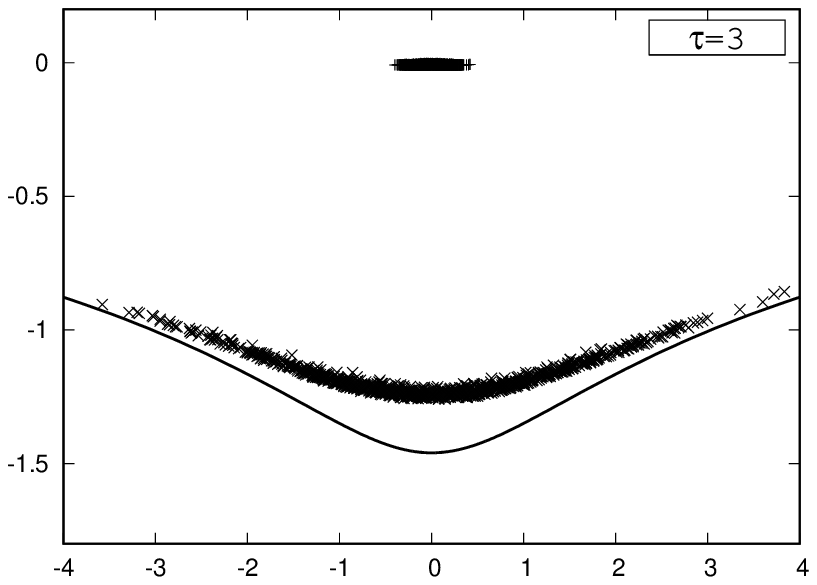}
\includegraphics[width=7cm]{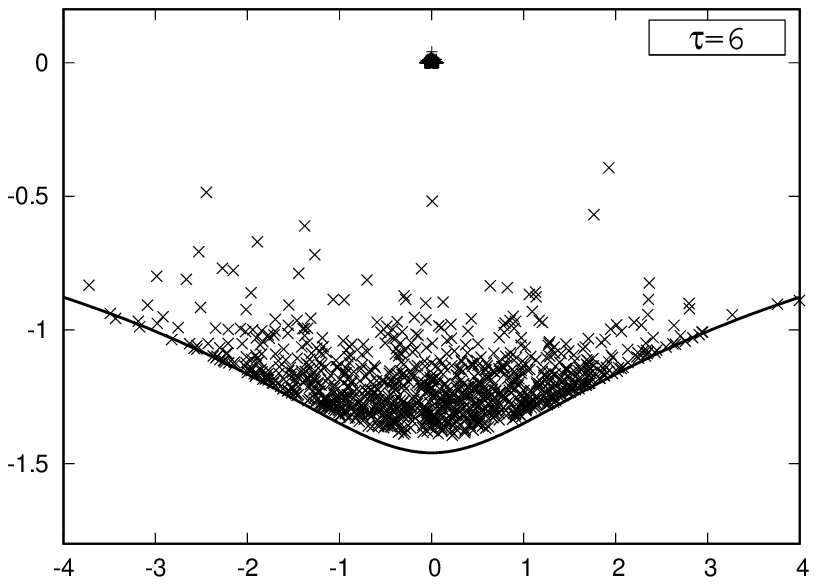}
\includegraphics[width=7cm]{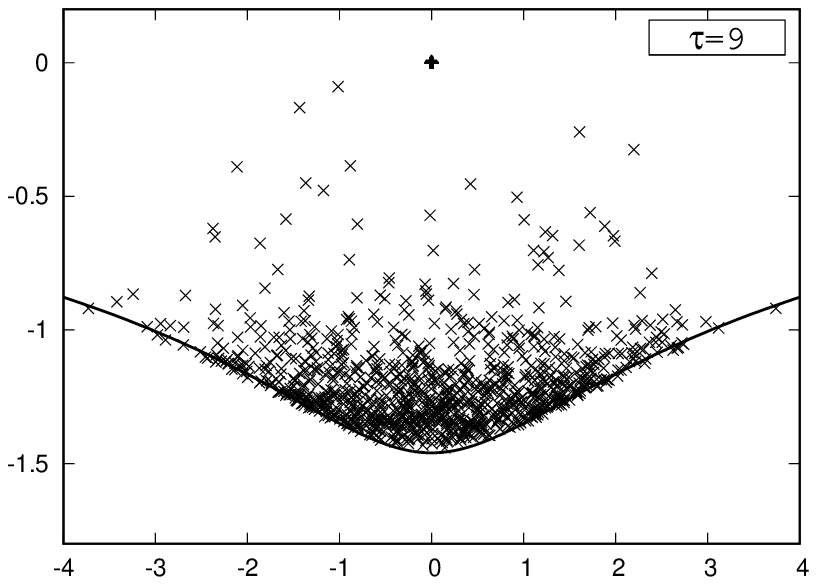}
\caption{The distribution of $z$ (represented by ``$+$'') and
flowed configurations (represented by ``$\times$'')
obtained by applying the CLM to \eqref{single variable Z deform} 
are plotted 
for $\tau=0,3,6,9$ from Top-Left to Bottom-Right.
The solid line represents the Lefschetz thimble.
At $\tau=0$, the distribution of $z$ and flowed configurations
coincides.}
\label{fig case1}
\end{figure}

\begin{figure}[t]
\centering
\includegraphics[width=7cm]{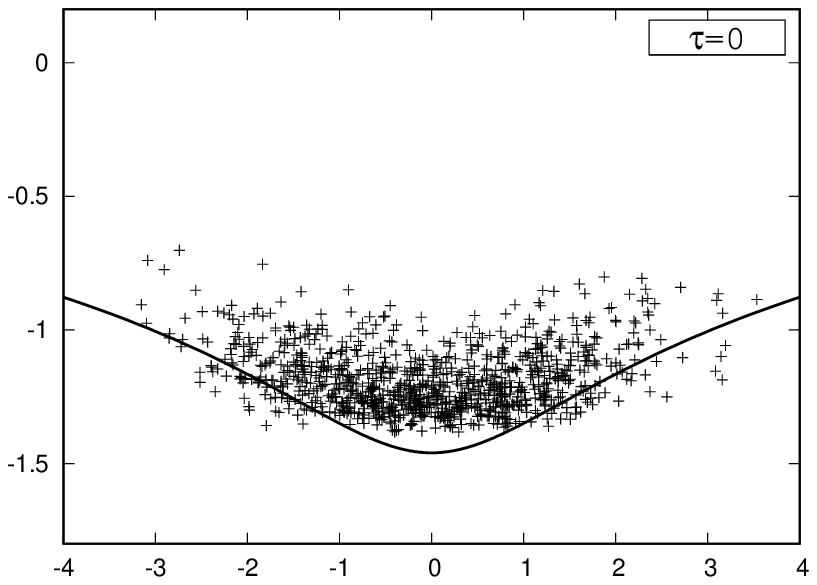}
\includegraphics[width=7cm]{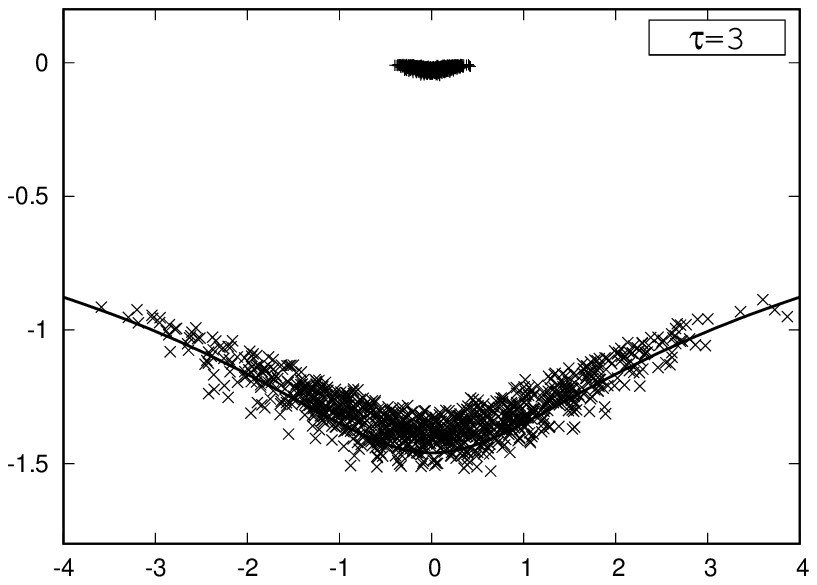}
\includegraphics[width=7cm]{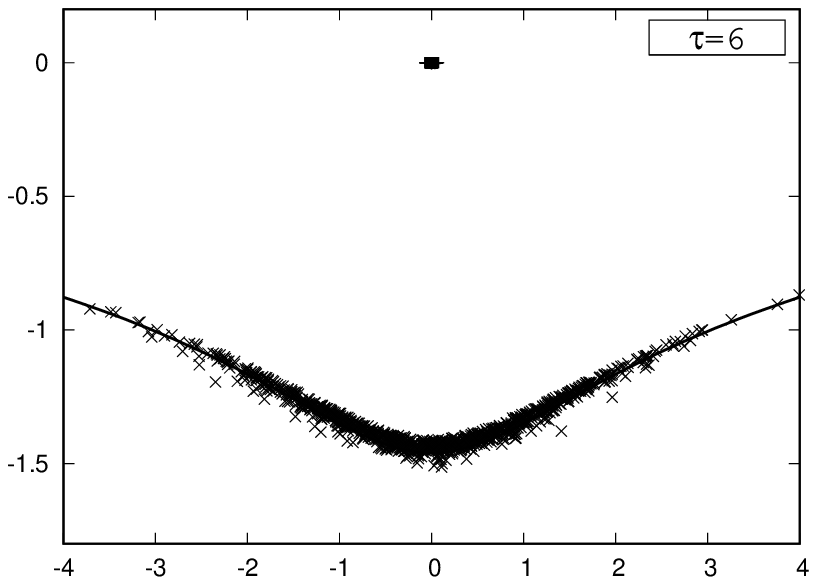}
\includegraphics[width=7cm]{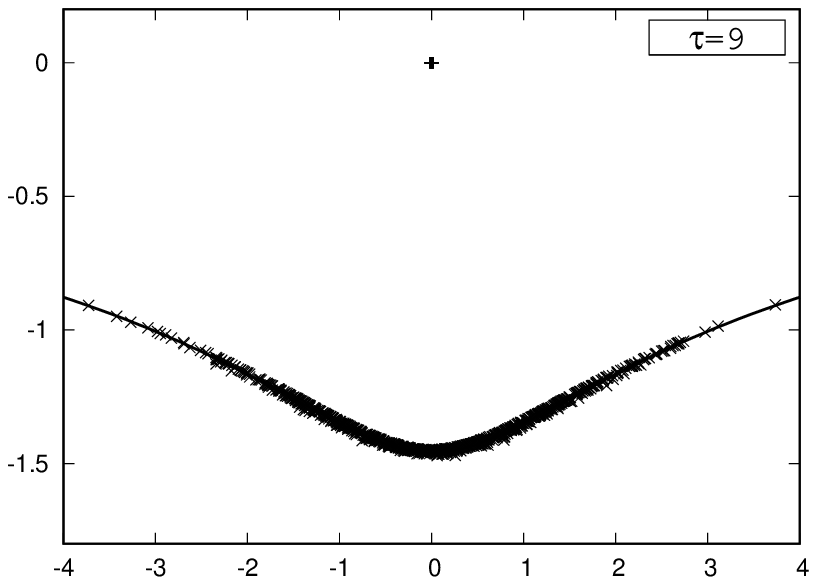}
\caption{The distribution of $z$ (represented by ``$+$'') and
flowed configurations (represented by ``$\times$'')
obtained by applying the CLM 
to the partially phase quenched model
(\ref{single variable Z deform-pPQ})
are plotted for $\tau=0,3,6,9$ from Top-Left to Bottom-Right.
The solid line represents the Lefschetz thimble.
At $\tau=0$, the distribution of $z$ and 
flowed configurations coincides.}
\label{fig case2}
\end{figure}

First, using the holomorphic gradient flow for a fixed flow time $\tau$, 
we rewrite \eqref{single variable Z} as
\begin{align}
Z=\int dx \,  J(x)  \,  (\phi(x)+i\alpha)^p \, e^{-\phi(x)^2/2} \ ,
\label{single variable Z deform}
\end{align}
where $\phi(x)=\phi(x;\tau)$ 
is obtained by solving \eqref{hge}
for $\phi(x;\sigma)$.
Here we consider three versions of our formulation; namely
\begin{enumerate}[label=(\roman*)]
\item CLM applied to the model (\ref{single variable Z deform}).
\item CLM applied to the partially phase-quenched model:
\begin{align}
Z_{\rm pPQ}=
\int dx \,  
|J(x)| \,  (\phi(x)+i\alpha)^p \, e^{-\phi(x)^2/2} \ .
\label{single variable Z deform-pPQ}
\end{align}
\item RLM applied to the totally phase-quenched model:
\begin{align}
Z_{\rm PQ}=
\int dx \,  
|J(x) \,  (\phi(x)+i\alpha)^p  \, e^{-\phi(x)^2/2} |  \ .
\label{single variable Z deform-PQ} 
\end{align}
\end{enumerate}
In the cases (ii) and (iii), appropriate reweighting is needed
in calculating the expectation value as described in the previous section.

\begin{figure}[t]
\centering
\includegraphics[width=7cm]{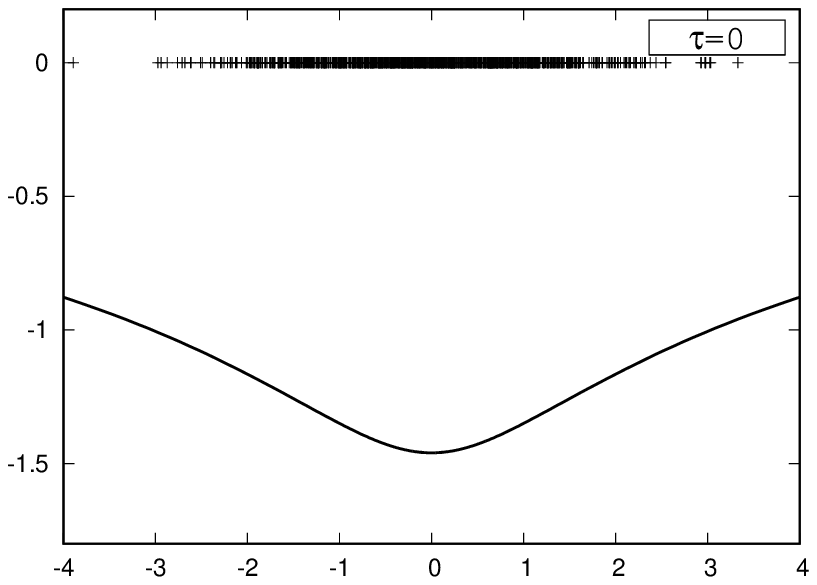}
\includegraphics[width=7cm]{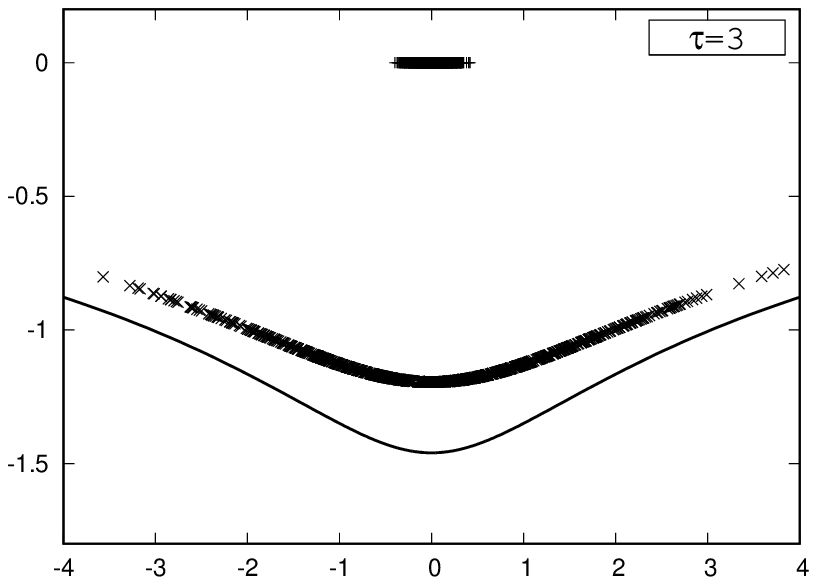}
\includegraphics[width=7cm]{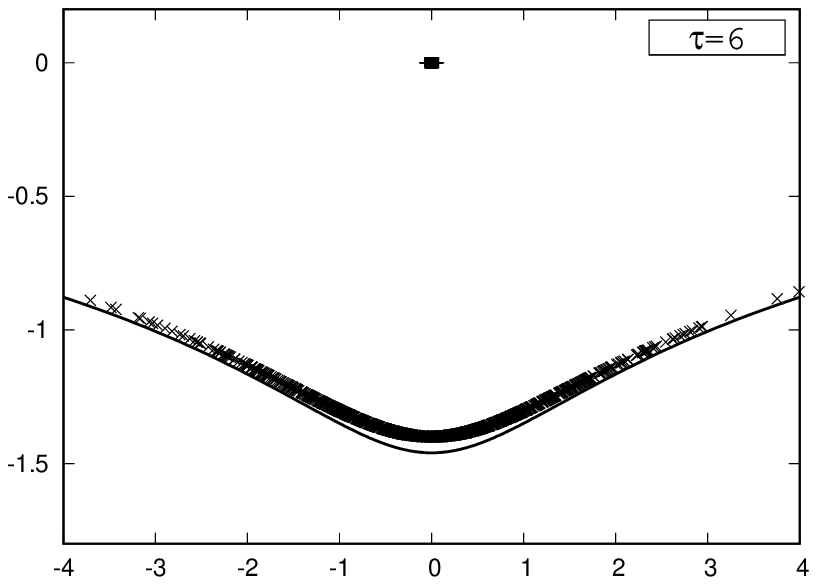}
\includegraphics[width=7cm]{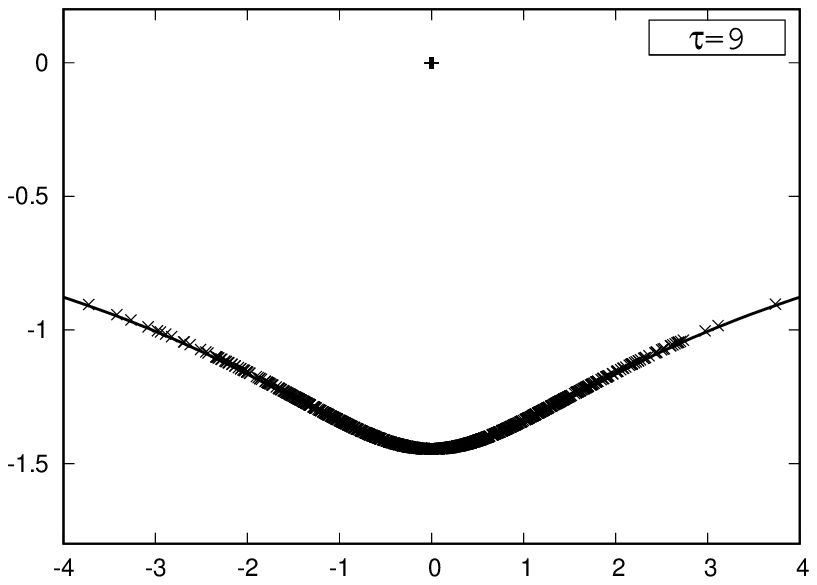}
\caption{The distribution of $z$ (represented by ``$+$'') and
flowed configurations (represented by ``$\times$'')
obtained by applying the RLM  to the phase quenched model
(\ref{single variable Z deform-PQ})
are plotted for $\tau=0,3,6,9$ from Top-Left to Bottom-Right.
The solid line represents the Lefschetz thimble.
At $\tau=0$, the distribution of $z$ and 
flowed configurations coincides.}
\label{fig case3}
\end{figure}

We focus on the choice of parameters $p=4$ and $\alpha=4.2$,
for which there exists only one Lefschetz thimble
and the ordinary CLM is known to be justified \cite{Nagata:2016vkn}.
The flow equation 
\eqref{hol hge-phi} and those for $J_{kl}(z;\sigma)$ and $K_{mlk}(z;\sigma)$
are solved for $\tau=0,3,6,9$.
For the details of the simulation, see ref.~\cite{Nishimura:2017vav}.
In figures~\ref{fig case1}, \ref{fig case2} and \ref{fig case3},
we show the distribution of $z$ and 
flowed configurations obtained 
in the case (i), (ii) and (iii), respectively. 


Let us first discuss the case (i) shown in figure~\ref{fig case1}.
At $\tau=0$, the distribution of $\phi(z;\tau=0)=z$ 
is nothing but that of the ordinary CLM.
At $\tau=3$, the distribution of $z$ comes close to the real axis,
which occurs because the sign problem is reduced to large extent
by the holomorphic gradient flow.
As a result of this,
the distribution of $\phi(z;\tau)$ comes close to a curve,
which looks similar to the one obtained in the GLTM for the same $\tau$
shown in figure~\ref{fig case3}.
At larger $\tau$, however,
the distribution of $\phi(z;\tau)$ spreads out again,
and its boundary approaches the Lefschetz thimble.
The spreading of the distribution
can be understood as the effects of
the complex Langevin dynamics,
which takes care of the residual sign problem arising 
from the Jacobian even in the large-$\tau$ limit.
The distribution of $z$ shrinks towards the origin,
but the zoom up shows that
it does not contract to a distribution on the real 
axis \cite{Nishimura:2017vav}.
It is interesting that
the expectation values evaluated
with flowed configurations $\phi(z;\tau)$ 
reproduce the exact results for any $\tau$ without reweighting
although the distribution of $\phi(z;\tau)$ looks quite different
for different $\tau$.


The case (ii) shown in figure~\ref{fig case2} is of particular interest.
At $\tau=0$, 
the distribution of $\phi(z;\tau=0)=z$ 
is nothing but that of the ordinary CLM, as in the case (i).
As $\tau$ increases, 
the distribution of $\phi(z;\tau)$ approaches the Lefschetz thimble,
while the distribution of $z$ approaches 
the real axis (See also the Appendix of ref.~\cite{Nishimura:2017vav}.).
This is due to the fact that the sign problem 
that the complex Langevin dynamics has to take care of
vanishes in the large $\tau$ limit
since the phase of the Jacobian is taken care of by reweighting
and the remaining phase factor
is taken care of by the holomorphic gradient flow.
Thus, in this case our unified formulation interpolates
the CLM and the original Lefschetz-thimble method via the flow time.

In the case (iii) shown in figure~\ref{fig case3},
our formulation reduces to the GLTM \cite{Alexandru:2015sua}.
At $\tau=0$, the calculation simply amounts to applying 
the reweighting method
to the original model (\ref{single variable Z}).
At $\tau > 0$, the distribution of $\phi(x;\tau)$ 
is restricted to the deformed integration contour $M_\tau$, which 
approaches the Lefschetz thimble as $\tau$ is increased.
Note also that the distribution of $x$ shrinks towards the origin 
as $\tau$ increases.


\section{Summary and discussions} 
\label{sec:summary}

We have proposed a formulation which unifies the CLM and the GLTM
by applying the CLM to the real variables 
that parametrize the deformed integration contour 
in the GLTM.
The residual sign problem in the GLTM is treated in three different
ways. 
We have applied the three versions
to a single-variable model in the case with a single Lefschetz thimble,
and investigated the distribution of flowed configurations,
which are used in evaluating the observables.
In particular, we have found that the second version 
interpolates the ordinary CLM
and the original Lefschetz-thimble method.

While our formulation is useful in clarifying the relationship
between the CLM and the GLTM, it is not of much practical use
since the computational cost required in solving
the differential equations for
$\phi(z;\sigma)$, $J_{kl}(z;\sigma)$ and $K_{mlk}(z;\sigma)$
increases as $O(V)$, $O(V^3)$ and $O(V^5)$, respectively,
with the system size $V$.
Note also that the computational cost increases linearly 
with the flow time. 
Furthermore, our formulation as it stands does not help
in enlarging the range of 
applicability of the CLM \cite{Aarts:2009uq,Aarts:2011ax,Nagata:2016vkn}.
The reason for this is that the distribution of flowed configurations
is qualitatively not much different from 
the distribution of configurations obtained in the CLM,
which is nothing but the distribution obtained for
$\tau=0$ in the first and second versions of our formulation.
For instance, in the model (\ref{single variable Z})
it is known that the CLM gives wrong results 
for $\alpha \lesssim 3.7$ with $p=4$ 
due to the singular drift problem \cite{Nagata:2016vkn}. 
This cannot be cured by 
the first and second versions of our formulation.
For large $\tau$, the situation gets even worse
since the flowed configurations tend to come closer 
to the singularity of the drift term,
and the simulation itself becomes unstable.
Thus, when the CLM fails, our formulation fails as well,
except for the third version, which does not rely on the justification
of the CLM.

Nevertheless, we consider that the relationship of the two methods as seen
in our unified formulation is useful 
in developing these methods further.
In particular, it is possible to generalize the CLM in such a way
that the range of applicability is enlarged as we report elsewhere.

\section*{Acknowledgement}
J.~N.\ was
supported in part by Grant-in-Aid for Scientific Research 
(No.~16H03988) from Japan Society for the Promotion of Science. 
S.~S.\ was supported by the MEXT-Supported Program 
for the Strategic Research Foundation at Private Universities 
``Topological Science'' (Grant No.~S1511006).

\bibliography{lattice2017}

\end{document}